\documentstyle[twoside,fleqn,espcrc2,epsfig]{article}

\def\beqn{\begin{eqnarray}}
\def\eeqn{\end{eqnarray}}
\def\bbbone{{\mathchoice {\rm 1\mskip-4mu l} {\rm 1\mskip-4mu l}
{\rm 1\mskip-4.5mu l} {\rm 1\mskip-5mu l}}}
\newcommand{\eq}[1]{(\ref{#1})}

\title{
\vspace{-9mm}
\rightline{\small ITEP-TH-29/99, KANAZAWA-99/08, UL-NTZ 19/1999}
\vspace{-2mm}
\rightline{\small 1 September, 1999}
Vortex profiles and vortex interactions at the electroweak
crossover\thanks{Presented by the first author at Lattice'99, Pisa, Italy.}
}
\author{
M.N. Chernodub\address{
Institute of Theoretical and Experimental
Physics, B.~Cheremushkinskaya 25, Moscow, 117259, Russia},
E.--M.~Ilgenfritz\address{Institute for Theoretical Physics,
Kanazawa University, Kanazawa 920-1192, Japan}
and A.~Schiller\address{Institut f\"ur Theoretische Physik
and NTZ, Universit\"at Leipzig, D-04109 Leipzig, Germany
}
}
\begin{document}
\begin{abstract}
Local correlations of Z--vortex operators with gauge and Higgs fields
(lattice quantum vortex profiles) as well as vortex two-point
functions are  studied in the crossover region near a  Higgs mass of $100$ GeV
within the $3D$ $SU(2)$ Higgs model.
The vortex profiles resemble certain
features of the classical vortex solutions in the continuum.
The vortex--vortex interactions are analogous to the
interactions of Abrikosov vortices in a type--I superconductor.
\end{abstract}
\maketitle

\section{Introduction}
\vspace{-1mm}
Although the standard model does not possess {\it topologically
stable} monopole-- and vortex--like defects, one can define so-called
{\it embedded} topological defects~\cite{VaBa69,BaVaBu94}: Nambu
monopoles~\cite{Na77} and $Z$--vortex strings~\cite{Na77,Ma83}.
Last year, we have started~\cite{ChGuIlSc98-1,ChGuIlSc98-2} to investigate
how the electroweak transition and the continuous crossover can be understood
in terms of the behavior of these excitations.
This has been done in the framework of dimensional reduction
which is reliable for Higgs boson masses between $30$ and $240$
GeV~\cite{generic}.
Due to the similarity of the phase transitions in the $SU(2)$ Higgs model
and in the $SU(2) \times U(1)$ electroweak theory \cite{SU2U1}, we
restricted ourselves to the $3D$ $SU(2)$ Higgs model.

In our lattice studies~\cite{ChGuIlSc98-1} we observed the vortices to
undergo a so--called percolation transition which coincides with the
first order phase transition at small Higgs masses. The percolation
transition continues to exist at realistic (large) Higgs
mass~\cite{ChGuIlSc98-2} when the electroweak theory has a smooth
crossover rather than a true thermal phase
transition~\cite{Kajantie}.
In our present study we have a closer look at the vortex properties
within the electroweak crossover regime [for a Higgs boson mass
$\approx 103$ GeV ($94$ GeV) in a $4D$ $SU(2)$ Higgs model with (without)
top quark].

\section{Lattice model and defect operators}
\vspace{-1mm}
To construct the vortices on the lattice we use their correspondence
to the Abrikosov-Nielsen-Olesen (ANO)
strings~\cite{ANO} embedded into an Abelian subgroup of the $SU(2)$ gauge
group. We define
a composite adjoint unit vector field $n_x = n^a_x \sigma^a$,
$n^a_x = - ({\phi^+_x \sigma^a \phi_x})/({\phi^+_x \phi_x})$, where
the 2--component  complex isospinor $  \phi_x$ is the Higgs field.
The field $n$ allows to  define the gauge invariant
flux ${\bar \theta}_p$ through the pla\-quet\-te $p=\{x,\mu\nu\}$ as
${\bar \theta}_p =  \arg ( {\mathrm {Tr}}
[(\bbbone + n_x) V_{x,\mu} V_{x +\hat\mu,\nu}
V^+_{x + \hat\nu,\mu} V^+_{x,\nu}])$ where
$V_{x,\mu}(U,n) =  U_{x,\mu} + n_x U_{x,\mu} n_{x + \hat\mu}$
is the projection of a link. Abelian link angles
$\chi_{x,\mu}=\arg\left(\phi^+_x V_{x,\mu} \phi_{x + \hat\mu}\right)$
are used to construct a pla\-quet\-te angle
$\chi_{p} = \chi_{x,\mu} + \chi_{x +\hat\mu,\nu} -
\chi_{x + \hat\nu,\mu} - \chi_{x,\nu}$.  The vorticity
$\sigma_p$ on the plaquette $p$ is~\cite{ChGuIl98}:
\beqn
\sigma_p = ( \chi_p - {\bar \theta}_p) \slash (2\pi) \,.
\label{SigmaNjN}
\eeqn
The vortex trajectories are formed by links ${}^* l=\{x,\rho\}$ of the
dual lattice (${}^* l$ dual to $p$) which carry a non--zero
vorticity ${}^* \sigma_{x,\rho} = \varepsilon_{\rho\mu\nu}
\sigma_{x,\mu\nu} \slash 2$. Those
trajectories are either closed or begin/end on Nambu (anti-)
monopoles.

\section{Vortex profiles}
\vspace{-1mm}
The topologically unstable vortices populate the ``vacuum'' of the model
due to thermal fluctuations.
Looking at realistic equilibrium configurations~\cite{EW98} we would not
expect to find vortices with classical (ANO type, Refs.~\cite{ANO}) profiles.
Our lattice vortex defect
operator \eq{SigmaNjN} is constructed to detect a line-like object (in
$3D$ space--time) with non-zero vorticity (``soul of the physical
vortex'').
Within a given gauge--Higgs configuration, the vortex profile around that
soul would by screened by quantum fluctuations, compared to its
classical shape.
However, the average over all vacuum vortices may reveal a structure
resembling a classical vortex to some extent. This requires to
study the correlators of the vortex operators with various field
operators.
Profiles obtained in this way are called ``quantum vortex
profiles'' although {\it thermal} fluctuations contribute to the correlators,
too.

Preliminary indications of some vortex-like structure have been given
in Ref.~\cite{ChGuIlSc98-2}. While in the
center of a classical {\it continuum} vortex the Higgs field modulus is zero
and the energy density reaches its maximum~\cite{Na77,BaVaBu94}, in thermal
equilibrium {\it on the lattice} the (squared) modulus of the Higgs
field, $\rho^2 = (\phi^+_x \phi_x )$ (the gauge field energy
density, $E^g = 1-\frac12 \mathrm{Tr} U_p$)
were found lower (higher) on the vortex trajectories than the
corresponding bulk averages\footnote{A similar
method was used to study the physical features of Abelian
monopoles in $SU(2)$ gluodynamics, Ref.~\cite{MIP}.}.

We define the following vortex--field correlators
for plaquettes $P_0$ and $P_R$ located in the same plane
(perpendicular to a local segment of the vortex trajectory)
\beqn
C_\rho(R) \!=  <\!\!\sigma^2_{P_0} \, \rho^2_{P_R}\!\!\!>\,,\,\,\,
C_E(R)    \!=  <\!\!\sigma^2_{P_0} \, E^g_{P_R}\!\!\!>\,,
\label{CorrDefs}
\eeqn
where $R$ is the distance between the plaquettes.
The Higgs modulus is $\rho^2_P = (1 \slash 4)\,
\sum_{x \in P} \rho^2_x$ (averaged over the corners of $P$).

\begin{figure}[!htb]
\vspace{2mm}
 \begin{minipage}{7.5cm}
 \begin{center}
  \epsfig{file=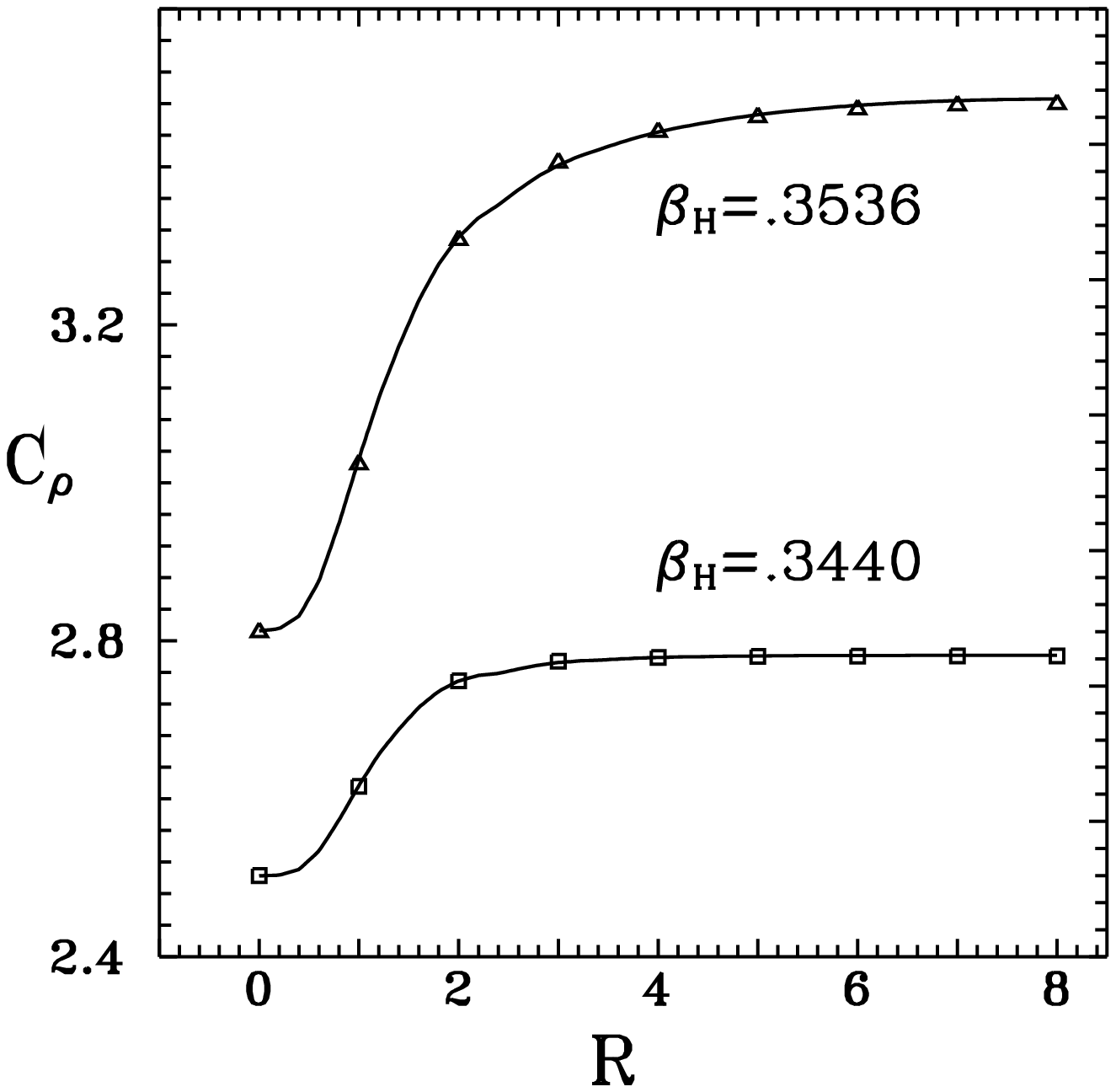,width=3.6cm,height=3.0cm}
  \epsfig{file=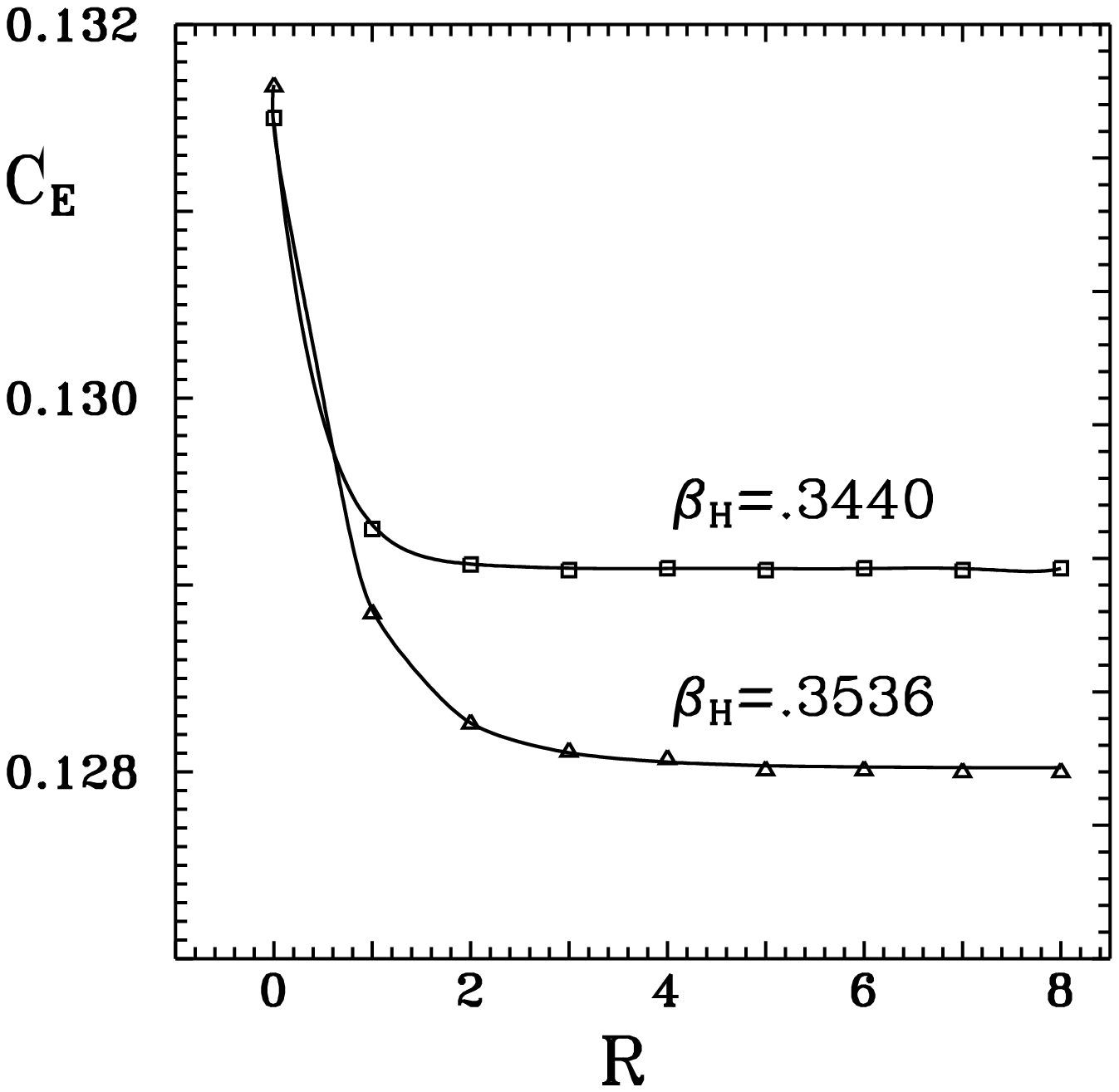,width=3.6cm,height=3.0cm}\\
(a) \hspace{3.0cm} (b)\\
 \end{center}
 \end{minipage}
\vspace{-8mm}
 \caption{\small Examples of vortex profiles for the Higgs field
(a) and gauge field energy (b). Solid lines represent fits \eq{fits}.}
\label{vortex profiles}
\vspace{-6mm}
\end{figure}

The quantum vortex profiles for the Higgs field, $C_\rho(R)$, and gauge
field energy, $C_E(R)$, are shown in Figures~\ref{vortex profiles}(a) and
\ref{vortex profiles}(b), respectively,
for two values of the hopping
parameter: $\beta_H=.3440$ (symmetric side) and $\beta_H=.3536$ (on top of
the crossover) for a lattice $16^3$ at gauge coupling\footnote{Results for
larger $\beta_G$ and bigger lattices will be presented elsewhere.}
$\beta_G=8$. On the Higgs side the profiles are qualitatively similar.
One can see from these Figures that the vortices have, on the average,
a thickness of several lattice spacings while the
center of the vortex is located within the plaquette
$P_0$ [identified by the vortex current \eq{SigmaNjN}].

To parametrize the vortex shape
we fit the correlator data \eq{CorrDefs} by the
following functions:
\begin{eqnarray}
\label{fits}
C^{\mathrm{fit}}_\rho(R)\!\! & = & \!\!
C_\rho + B_\rho  \!\!\! \sum\limits_{x\in P_0, \, y\in P_R} \!\!\!
G(\vert {\mathbf x} - {\mathbf y}\vert; m_\rho)\,,
\nonumber \\
C^{\mathrm{fit}}_E(R)\!\! & = &  \!\!C_E + B_E \, G(R; m_E)
\end{eqnarray}
with asymptotic values $C_{\rho,E}$ as well as  amplitudes $B_{\rho,E}$
and
inverse coherence lengths (effective masses) $m_E$ and $m_\rho$.
The {\it lattice} function $G(R; m)$ was
proposed to fit {\it point--point} correlation
functions in Ref.~\cite{Engels:1995ek}.
The function $G$ is proportional to the scalar
propagator with the mass $2 \sinh (m/2)$ in $3D$ space.
Best fits are shown in Figures \ref{vortex profiles} by solid lines.

\begin{figure}[!htb]
\vspace{-6mm}
 \begin{minipage}{7.5cm}
 \begin{center}
  \epsfig{file=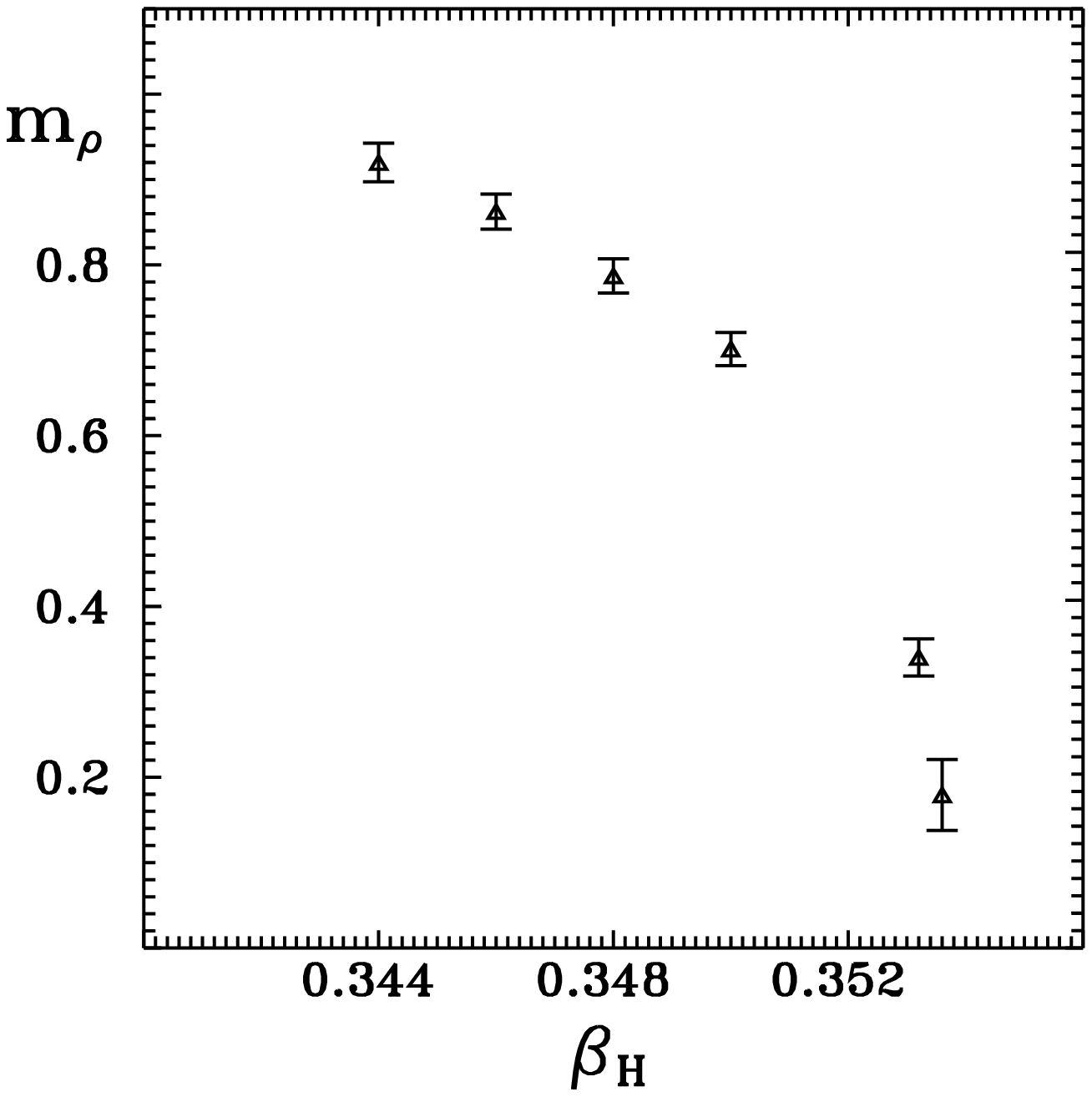,width=3.6cm,height=3.0cm}
  \epsfig{file=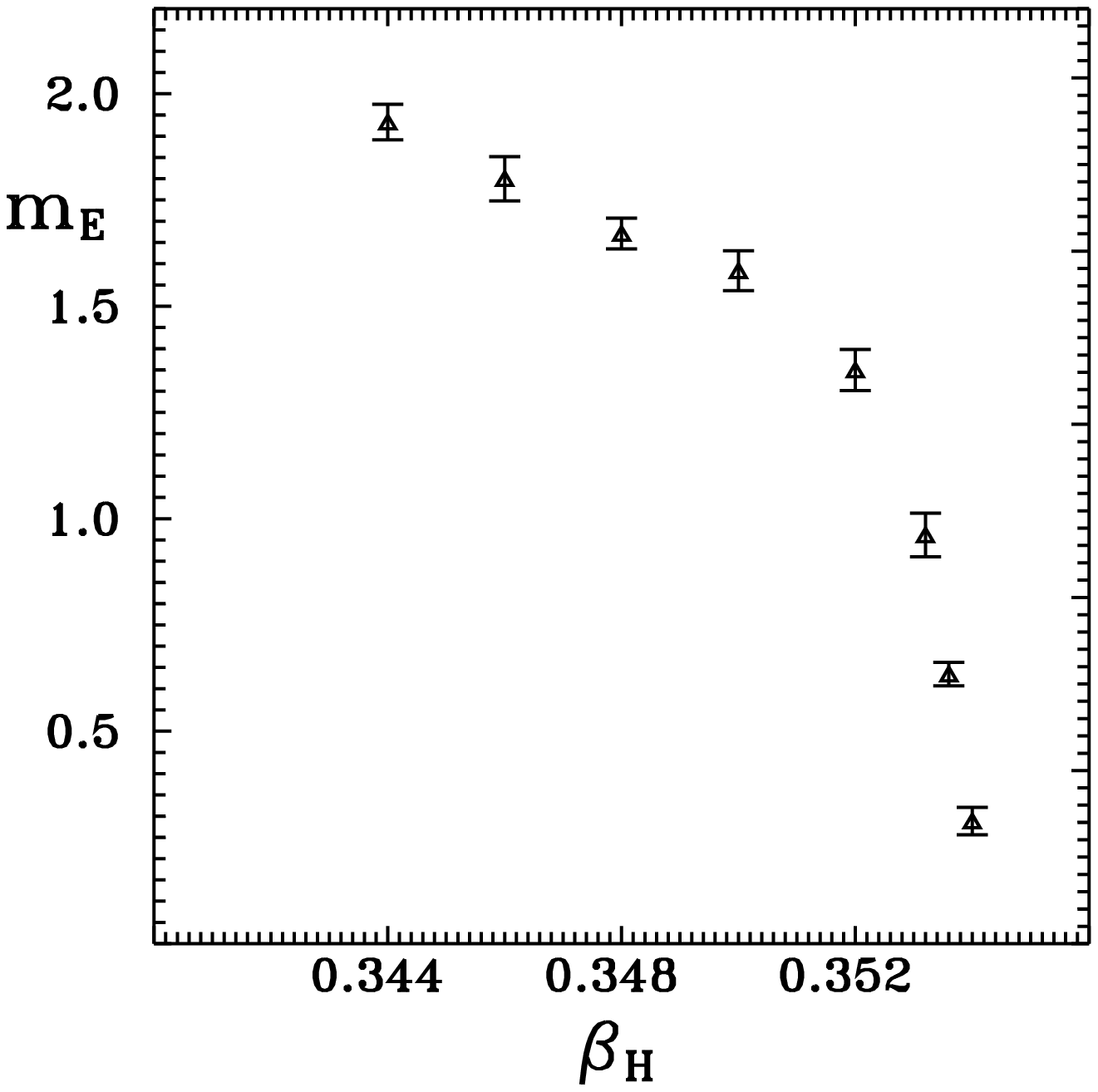,width=3.6cm,height=3.0cm}\\
  (a) \hspace{3.0cm} (b) \\
 \end{center}
 \end{minipage}
\vspace{-8mm}
 \caption{\small Fitted mass parameters of the quantum vortex
Higgs (a) and energy (b) profiles.}
 \label{masses}
\vspace{-6mm}
\end{figure}

As an example, the effective masses as function of the hopping parameter
(decreasing temperature) are shown in
Figures~\ref{masses}
on the symmetric side of the crossover.
The masses become
small near the crossover (where the fit ceases to be good). Deeper on the
symmetric side, the quantum vortex profiles are squeezed compared to the
classical ones due to Debye screening leading to a  smaller coherence length.
Approaching the crossover from the symmetric side the
density of the vortices becomes smaller reducing this effect.

\vspace{-2mm}
\section{Type of vortex vacuum}
\vspace{-1mm}
A vortex medium can be characterized in superconductor terms: if
two static vortices with the same vorticity attract (repel) each
other the medium corresponds to a type--I (type--II)
superconductor. In order to define
the type of interaction for the case of
electroweak matter we have measured two--point functions
of the vortex currents:
\beqn
C_+(R) \!\! & = & \!\! < \!\!|\sigma_{P_0}|\,|\sigma_{P_R}| \!\!>
=  2 (g_{++} + g_{+-}) \nonumber\\
C_-(R) \!\! & = & \!\! < \!\!\sigma_{P_0} \,\sigma_{P_R} \!\!>
=  2 (g_{++} - g_{+-})
\label{jj}
\eeqn
where $g_{+\pm}=g_{+\pm}(R)$ is shorthand for contributions to the
correlation functions $C_\pm$ of parallel or anti--parallel vortices at
positions $P_0$ and $P_R$. Obviously, $g_{++}=g_{--}$ and $g_{+-}=g_{-+}$.

The correlators $g_{++}(R) = (C_+ + C_-) \slash 4$ ($g_{+-}(R) = (C_+ -
C_-) \slash 4$) can be interpreted as the
average density of vortices (anti-vortices)
relative to the bulk density
(normalized to unity at $R \to \infty$)
in the plane orthogonal to a vortex current at distance $R$.

If the vortices attract/repel each other (type--I/type--II) the long
range tail of the function $g_{++}(R)$
should exponentially approach unity from above/below,
while the behavior of $g_{+-}(R)$ is attractive independent of the type
of superconductivity.

\begin{figure}[!htb]
\vspace{-10mm}
 \begin{center}
  \epsfig{file=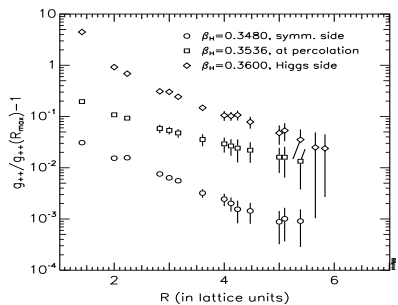,width=5.2cm,height=4.0cm}
\vspace{-10mm}
\caption{\small Normalized correlator $g_{++}(R)$.}
\label{type}
\vspace{-5mm}
\end{center}
\end{figure}

In Figure~\ref{type}
 one can see that the same-vorticity pair distribution
$g_{++}$\footnote{Shortest distances and data points of the order of the
noise have been omitted.}
decreases exponentially with a slope that becomes minimal on the
crossover. Apart from $R<2$ the opposite-vorticity distribution $g_{+-}$
behaves similarly.
Thus we conclude that electroweak matter at the crossover belongs to the
type--I vortex vacuum class.
The attractive character becomes even stronger on the
Higgs (lower temperature) side.

\section*{Acknowledgments}
\vspace{-1mm}
M.N.~Ch. was partially supported by grants INTAS-RFBR-95-0681,
RFBR-99-01230a and INTAS 96-370.


\vspace{-2mm}

\begin{thebibliography}{99}

\bibitem{VaBa69}
T.~Vachaspati and M.~Barriola, {\it Phys. Rev. Lett.}
{\bf 69} (1992) 1867.

\bibitem{BaVaBu94} M.~Barriola, T.~Vachaspati, M.~Bucher,
{\it Phys.~Rev.} {\bf D50} (1994) 2819.

\bibitem{Na77}
Y.~Nambu, {\it Nucl.~Phys.} {\bf B130} (1977) 505.

\bibitem{Ma83}
N.S.~Manton, {\it Phys.~Rev.} {\bf D28} (1983) 2019.

\bibitem{ChGuIlSc98-1}
M.N.~Chernodub, F.V.~Gubarev, E.--M.~Ilgenfritz, A.~Schiller,
{\it Phys.~Lett.} {\bf B434} (1998) 83.

\bibitem{ChGuIlSc98-2}
M.N.~Chernodub, F.V.~Gubarev, E.--M.~Ilgenfritz, A.~Schiller,
{\it Phys.~Lett.} {\bf 443} (1998) 244.

\bibitem{generic} K.~Kajantie et al., {\it Nucl. Phys.} {\bf
B458} (1996) 90;
M.~G\"urtler et al., {\it ibid.} {\bf B483} (1997) 383.

\bibitem{SU2U1}
K.~Kajantie, M.~Laine, K.~Rummukainen, M.~Shaposhnikov,
{\it Nucl. Phys.} {\bf B493} (1997) 413.

\bibitem{Kajantie}
K.~Kajantie {\it et al}, {\it Phys. Rev. Lett.} {\bf 77}
(1996) 2887;
M.~G\"urtler, E.--M.~Ilgenfritz, A.~Schiller,
{\it Phys.~Rev.} {\bf D56} (1997) 3888.

\bibitem{ANO}
A.A.~Abrikosov, {\it Sov. Phys. JETP} {\bf 32} 1442 (1957);
H.B.~Nielsen and P.~Olesen, {\it Nucl.~Phys.} {\bf 61} (1973) 45.

\bibitem{ChGuIl98}
M.N.~Chernodub, F.V.~Gubarev, E.--M.~Ilgenfritz,
{\it Phys.~Lett.} {\bf B424} (1998) 106.

\bibitem{EW98}
M.N.~Chernodub, F.V.~Gubarev, E.--M.~Ilgenfritz, A.~Schiller,
{\tt hep-ph/9902285}.

\bibitem{MIP} B.L.G.~Bakker, M.N.~Chernodub, M.I.~Polikarpov,
{\it Phys.~Rev.~Lett.} {\bf 80} (1998) 30.

\bibitem{Engels:1995ek} J.~Engels, V.K.~Mitryushkin, T.~Neuhaus,
{\it Nucl.~Phys.} {\bf B440} (1995) 555.

\end{thebibliography}
\end{document}